\documentclass[journal,twoside]{IEEEtran}
\usepackage{amssymb}
\usepackage{amsthm}
\usepackage{multirow}
\usepackage{inputenc}
\usepackage{enumerate} 
\usepackage{textcomp}
\usepackage{listings}
\usepackage{array}
\usepackage[switch,mathlines]{lineno}
\usepackage{soul}
\usepackage{xfrac}
\usepackage{graphicx}
\usepackage{cite}
\usepackage[cmex10]{amsmath}
\usepackage{mathrsfs}
\usepackage{algorithm}
\usepackage{xcolor}
\usepackage{colortbl}
\usepackage{makecell}
\usepackage{cancel}
\usepackage{algorithm,algpseudocode}% http://ctan.org/pkg/{algorithms,algorithmx}
\algnewcommand{\Inputs}[1]{%
  \State \textbf{Inputs:}
  \Statex \hspace*{\algorithmicindent}\parbox[t]{.8\linewidth}{\raggedright #1}
}
\algnewcommand{\Initialize}[1]{%
  \State \textbf{Initialize:}
  \Statex \hspace*{\algorithmicindent}\parbox[t]{.8\linewidth}{\raggedright #1}
}
\makeatletter
\def\footnoterule{\kern-3\p@
  \hrule \@width 3.3in \kern 2.6\p@} % the \hrule is .4pt high
\makeatother

\usepackage{algpseudocode}

\usepackage{stackengine}

\usepackage{algpseudocode}
\usepackage{CJK}
\usepackage[cmex10]{amsmath}
\usepackage{bm}
\usepackage{color}
\usepackage{amsmath,amsthm,amssymb,amsfonts}
\usepackage{flushend}
\usepackage{float}

\newtheorem{remark}{Remark}

\usepackage{arydshln} %to include dashed lines in arrays
\usepackage{hyperref}
\hypersetup{
     colorlinks   = true,
     linkcolor    = blue,
     citecolor    = red,
     urlcolor     = blue
}
\makeatletter
\newcommand*{\transpose}{%
  {\mathpalette\@transpose{}}%
}
\newcommand*{\@transpose}[2]{%
  \raisebox{\depth}{$\m@th#1\intercal$}%
}
\makeatother

\usepackage{soul}
\usepackage{tikz}
\usepackage{booktabs}% http://ctan.org/pkg/booktabs
\usepackage{tabularx}% http://ctan.org/pkg/tabularx

\usepackage[caption=false,font=footnotesize]{subfig}
\ifCLASSINFOpdf
\else
\fi

\begin{document}
\renewcommand{\ttdefault}{cmtt}
\bstctlcite{IEEEexample:BSTcontrol}

\title{Preconditioner-Based Acceleration Method for Solving EMTP Linear Equations
}
%: A Regularization Method for EMTP Problems
% Decoupled Jacobi-preconditioner for Dominant Transient Problem

% <-this % stops a space

% <-this % stops a space

\author{
{Qi Lou,~\IEEEmembership{Student Member},
Yijun~Xu,~\IEEEmembership{Senior Member}, 
Yang Cao,~\IEEEmembership{Student Member},
%~\IEEEmembership{Student Member},
Wei Gu,~\IEEEmembership{Senior Member},
Fei Zhang,~\IEEEmembership{Member}, 
}
\vspace{-0.5cm}
%~\IEEEmembership{Student Member},
%       \thanks{Y. Li, Y. Xu, W. Gu, and  S. Lu,  are with the  Electrical Engineering Department, Southeast University, Nanjing, Jiangsu 210096, China
       %, (e-mail:\{yiboli, yijunxu,  wgu\}@seu.edu.cn)
 %      .}% <-this % stops a space
%\thanks{L. Mili is with the Bradley Department of Electrical and Computer Engineering, Virginia Tech,
%Falls Church, VA 22043， USA 
%(e-mail:\{lmili\}@vt.edu)
%.}
%~\IEEEmembership{Student Member},
\thanks{
Q. Lou, Y. Xu, Y Cao, F. Zhang and W. Gu are with the  Electrical Engineering Department, Southeast University, Nanjing, Jiangsu 210096, China (e-mail: \{\texttt{louqi, yijunxu, yangcao, zhangfei, wgu\}@seu.edu.cn}).
\\\emph{(Corresponding author: Wei Gu.)}
}
%\thanks{S. Yao is with the School of Engineering, Cardiff University, UK, CF24 3AA (e-mail: \texttt{YaoS8@cardiff.ac.uk}).}
%\thanks{L. Mili is with the Bradley Department of Electrical and Computer Engineering, Virginia Tech, Northern Virginia Center, Falls Church, VA 22043 USA (e-mail: \texttt{lmili@vt.edu}).}
%\thanks{M. Netto is with the Power Systems Engineering Center, National Renewable Energy Laboratory, Golden, CO 80401, USA (e-mail:\{marcos\}@nrel.gov).}
% <-this % stops a space
%\thanks{M.~Korkali is with the University of Missouri, Department of Electrical Engineering and Computer Science, Columbia, MO 65211 USA (e-mail: \texttt{korkalim@missouri.edu}).}
%\thanks{The study was supported by the science and techeating networkology project of the State system Corporation of China, Grant 5108-202218280A-2-366-XG. %\emph{(Corresponding author: Yijun Xu.)}
%}
}

%\markboth{IEEE Transactions on Industrial Informatics}%
%{Li \MakeLowercase{\textit{\textit{et al.}}}: Generalized Unscented Transformation for Variance-based Sensitivity Analysis for IES}\
\markboth{CSEE Journal of Power and Energy Systems}%
{Lou \MakeLowercase{\textit{\textit{et al.}}}: Preconditioner-Based Acceleration Method for Solving EMTP Linear Equations}

\maketitle
% \vspace{-0.5cm}
\begin{abstract}
The computational speed of electromagnetic transient programs (EMTP) is severely limited by both the curse of dimensionality and the ill-conditioned system matrix, which collectively degrade solver performance. However, existing research on EMTP acceleration has largely overlooked the issue of ill-conditioning. {This letter presents a first systematic, EMT-oriented investigation of the ill-conditioning of the EMTP admittance matrix by establishing a link between its physical origins and mathematical pathologies, thereby revealing the underlying mechanism by which network topology induces ill-conditioning.} Building upon these structural insights, a preconditioner-based strategy is developed that significantly accelerates computation while preserving numerical accuracy. Simulation results demonstrate the outstanding efficiency and robustness of the proposed approach.

\end{abstract}


\IEEEpeerreviewmaketitle

\vspace{-0.3cm}
\section{Introduction}
\IEEEPARstart{C}{o}mputational efficiency remains the central challenge in electromagnetic transient (EMT) simulations \cite{CSEE1}. While significant efforts have been devoted to developing improved linear solvers, including the Conjugate Gradient (CG)\cite{CG} and Minimal Residual (MR) methods\cite{MR}, {solver-level acceleration alone has not fully resolved this challenge.} The fundamental computational requirements - $\mathcal{O}(n)$ for CG and $\mathcal{O}(\text{nnz} \cdot m)$\footnote{Here, $\text{nnz}$ is the number of non-zero matrix entries, and $m$ is the dimension of the Krylov subspace.} for MR - continue to present critical limitations that hinder complete solution of this persistent challenge\cite{cg_mires}.

{From a broader perspective, EMT acceleration has been pursued through multiple complementary directions. Most existing efforts focus on physical modeling or implementation aspects, such as network decoupling\cite{EMT_decouple}, model simplification\cite{EMT_model}, and high-performance computing\cite{EMT_GPU,EMT_FPGA}. Solver-oriented acceleration from the mathematical perspective, which can complement these approaches, has received limited attention in EMT simulations.}

To further enhance EMTP computing efficiency from this solver-oriented perspective, this letter investigates the ill-conditioning of EMTP nodal equations. This focus is motivated by the fact that the condition number $\kappa$ directly governs numerical stability and computational complexity: for example, the Krylov subspace dimension in MR methods grows linearly with $\kappa$, while CG exhibits a complexity of $\mathcal{O}(\sqrt{\kappa})$~\cite{subdim}.

Motivated by this, this letter tends to accelerate the EMTP by pioneering an investigation into the ill-conditioning characteristics of EMT simulations through the development of tailored preconditioning techniques. {Although preconditioning has been reported in long-term time-domain simulations\cite{TPS1}, existing applications are mainly restricted to systems without power electronic devices and rely on empirical performance comparisons, lacking a systematic physical interpretation for preconditioner design in EMTP nodal equations. By explicitly exploiting the structural and physical characteristics of EMT modeling,} our method ensures numerically stable transformations while maintaining solution accuracy, enabling efficient resolution of EMTP.

\section{PRELIMINARIES of EMTP}

EMTP analyzes power system transients using discretization methods and nodal analysis \cite{EMTPBOOK}. Its matrix form is
\begin{equation}\label{EMTP}
     \mathbf{G}\vec{v}(t) = \vec{i}_{inj}(t)+\vec{I}_{history}\equiv \vec{i}(t).
\end{equation}
Here, $\Vec{v}(t)$ denotes the nodal voltages, $\vec{i}(t)$ denotes the nodal injected current, $\vec{i}_{inj}(t)$ denotes the externally injected current, $\Vec{I}_{history}$ denotes historical nodal injected current, and $\mathbf{G}$ denotes the network admittance matrix.
% , the specific construction process is
% \begin{equation}\label{G_mat}
%     g_{ij=}\begin{cases}
% -Y_{ij}&, i \neq j
%  \\
% Y_{i0}+\sum_{j \neq i}Y_{ij} &,i = j
% \end{cases}
% \end{equation}
% Here, $g_{ij}$ represents the element in the $i$-th row and $j$-th column, $Y_{ij}$ denotes the branch admittance between node $i$ and node $j$, and $Y_{i0}$ represents the shunt admittance from node $i$ to ground. 
Iterative solutions of \eqref{EMTP} yield the nodal voltages at each time step. To effectively mitigate the ill-conditioning of this solver, it is crucial to first examine its root causes from a physical perspective.

\vspace{-0.3cm}
\section{Analysis of the Causes of Ill-Conditioning}
% or rounding errors during computation
First, define the condition number $\kappa$ of the EMTP admittance matrix $\mathbf{G}$ as:
\begin{equation}\label{cond_num}
    \kappa(\mathbf{G})=\|\mathbf{G}\|_2\cdot\|\mathbf{G}^{-1}\|_2=\frac{\lambda_{max}(\mathbf{G})}{\lambda_{min}({\mathbf{G}})}
\end{equation}
where $\|\cdot\|_2$ represents the 2-norm of the matrix, $\lambda_{max}$ and $\lambda_{min}$ denote the largest and smallest eigenvalues of $\mathbf{G}$, respectively. 

In EMTP, $\mathbf{G}$ is typically ill-conditioned, exhibiting extreme sensitivity to small perturbations. For example, modeling a converter with a binary resistive switching device yields a condition number of about $7.998\times10^{12}$, whereas the standard IEEE 13-bus network without converters shows $3.345\times10^{7}$. Both values far exceed the threshold for well-conditioned matrices (typically $\kappa < 10^3$
). Using the physical structure of EMT networks, we first analyze the cause of ill-conditioning in EMTP:

\subsubsection{Grounded admittances}
In EMT networks, power sources or capacitors with large equivalent admittances are frequently grounded. Such large admittances make the corresponding node voltages highly sensitive to current injections, leading to ill-conditioned system matrices.
% Hence, the excessively large diagonal entries in these deviated rows are identified as a key contributor to the ill-conditioning of the admittance matrix.

\subsubsection{Switch Devices}
% Switches, modeled as binary resistors, exhibit extremely high admittance when closed. This leads to deviated rows in the admittance matrix dominated by switch admittance, thereby invalidating the condition in \eqref{gmm_chara}. Consequently, ill-conditioning in these cases requires reexamination from a physical perspective.

When a switch is closed, Kirchhoff’s voltage and current laws enforce nearly identical voltages and balanced currents at the connected nodes. This results in a strong coupling of voltage and current between the two nodes, leading to severe ill-conditioning.
Then, the admittance of closed switches is significantly higher than that of grounding elements, making switches the \textbf{primary cause}. 

Next, we analyze how these factors manifest themselves as ill-conditioned matrix components, facilitating our proposed effective conditioning improvement methods.
\vspace{-0.2cm}
\section{The proposed preconditioned EMTP equation-solving framework}
% \vspace{-0.1cm}
To address matrix ill-conditioning, we propose a preconditioned EMTP equation-solving framework for the first time. If a nonsingular matrix $\mathbf{M}$ exists such that it is approximately equal to $\mathbf{G}$, the preconditioned EMTP equation
\begin{equation}\label{pEMTP}
    \mathbf{M}^{-1}\mathbf{G}\vec{v}(t)= \mathbf{M}^{-1}\vec{i}(t)
\end{equation}
yields the same solution as the original system while ensuring a well-conditioned matrix. Here $\mathbf{M}$ is the preconditioner.
Designing an effective $\mathbf{M}$ requires fully capturing all ill-conditioning factors. Hence, a thorough analysis of how these factors affect the structure and properties of $\mathbf{G}$ is essential.
\subsubsection{Mathematical Analysis of Ill-Conditioning Factors}
The ill-conditioning in EMT simulations can be mathematically characterized using the Gershgorin Circle Theorem
\cite{Gershgorin_circle}. It states that all eigenvalues $\lambda$ of $\mathbf{G}$ lie within the union of $n$ Gershgorin discs, defined as 
\begin{equation}\label{origin_Gcircle}
    \bigcup_{i=1}^{n}\left\{\lambda_i \in \mathbb{C}:\left|\lambda_i-g_{i i}\right| \leq \sum_{j \neq i}\left|g_{i j}\right|\right\}
\end{equation}
Here, $g_{ij}$ represents the element in the $i$-th row and $j$-th column. Considering the real symmetric nature and diagonal dominance property of the EMTP admittance matrix, we have
\begin{equation}\label{Gershgorin_Circle}
    0<g_{ii}-\sum_{j \neq i}\left|g_{i j}\right|\leq\lambda_i \leq g_{ii}+\sum_{j \neq i}\left|g_{i j}\right|\leq2g_{ii}.
\end{equation}
Here, although \eqref{Gershgorin_Circle} is defined over the real domain, it remains applicable because the EMTP algorithm discretizes nonlinear elements into equivalent resistive forms\cite{CSEE2}, resulting in a real-valued admittance matrix.

Now, we can analyze the ill-conditioning introduced by grounded admittances. Since the branch admittances of nonswitched components are significantly smaller, \eqref{Gershgorin_Circle} implies that $\lambda_i \approx g_{ii}$, indicating that this form of ill-conditioning is primarily reflected in the diagonal entries. In contrast, ill-conditioning arising from closed switches is characterized by abnormally large diagonal and off-diagonal elements within certain rows, leading to near-linear dependencies among them. Therefore, $\mathbf{M}$ must account for both diagonal entries and the structural coupling introduced by closed switch branches.

Up to this point, most potential sources of ill-conditioning in EMTP matrices have been systematically analyzed. The analysis can be focused on fundamental circuit components for two primary reasons. {First, in EMTP simulations, the admittance matrix $\mathbf{G}$ is assembled from the discretized electrical network; control systems are evaluated separately and affect $\mathbf{G}$ only through changes in switching states or equivalent parameters, rather than through explicit control terms.} Second, even complex devices such as modular multilevel converters (MMC) are modeled as equivalent circuits of RLC components and switches \cite{MMCmodel}. Thus, analyzing RLC circuits and switching behavior is sufficient to capture the primary sources of ill-conditioning in typical simulations.
\begin{figure}[!htbp]%使用figure环境
\vspace{-0.3cm}
  \begin{center}
    \includegraphics[scale=0.4]{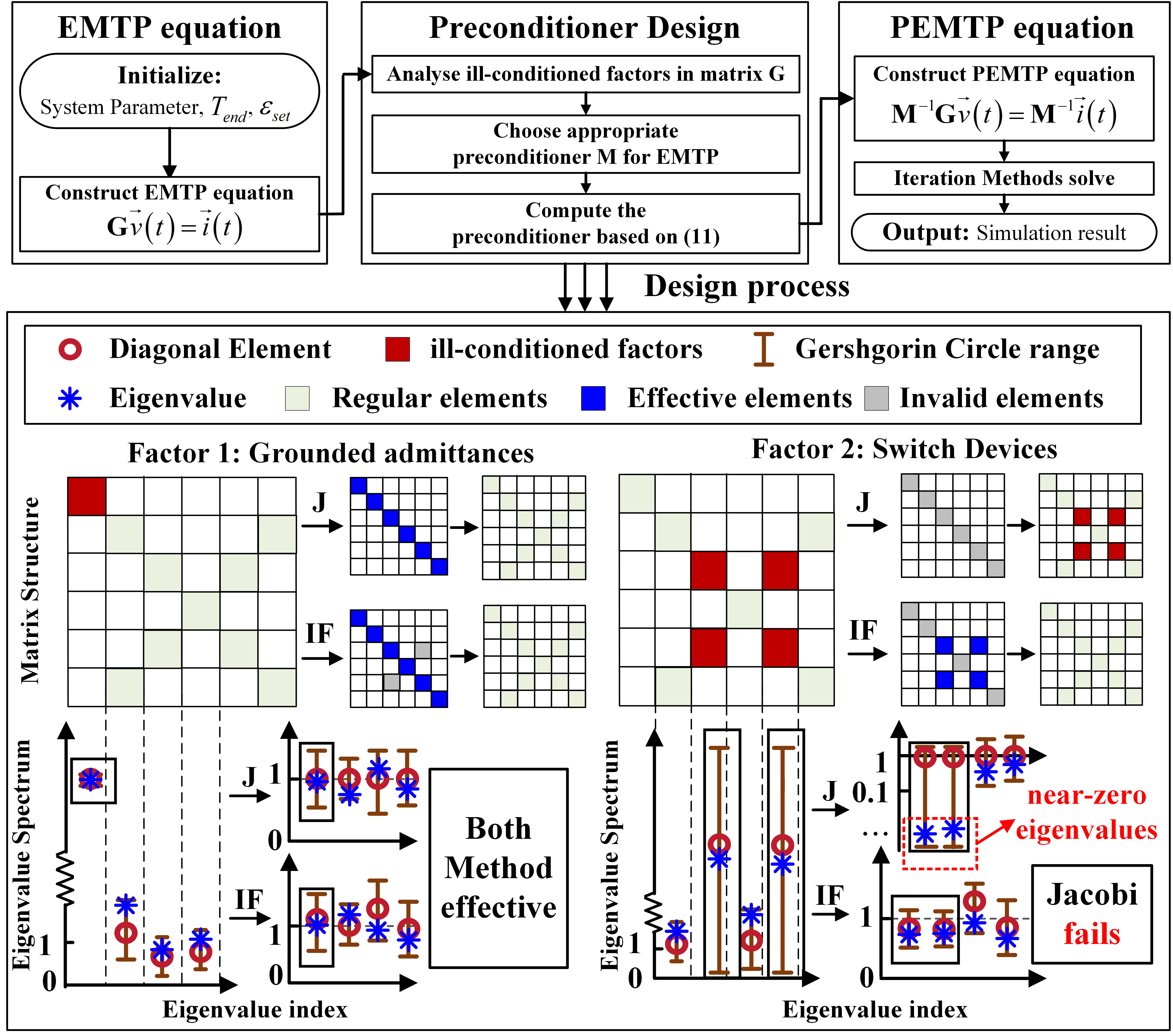}
    \setlength{\abovecaptionskip}{-0pt}
    \vspace{-0.3cm}
    \caption{PEMTP Solution Framework. In the workflow, the preconditioner design serves as a critical link between EMTP formulation and iterative solution. For factor 1, large diagonal entries lead to a dispersed eigenvalue spectrum, which Jacobi mitigates via row normalization that concentrates the spectrum. As a result, both Jacobi and IF provide effective preconditioning. In contrast, Factor 2 introduces structurally paired elements that cause near-linear row dependencies, reflected as broad Gershgorin circles. While Jacobi normalizes the diagonal, it fails to eliminate these dependencies or suppress near-zero eigenvalues (visible on a logarithmic scale), making it ineffective. In contrast, IF preserves the matrix structure and produces a more compact and centered eigenvalue spectrum, thus offering superior preconditioning.
}%添加标题
    \vspace{-0.5cm}
    \label{fig:precond}
  \end{center}
  
\end{figure}

\subsubsection{Jacobi Preconditioning}
First, considering that the ill-conditioning factors appear in diagonal entries, we first attempt the Jacobi preconditioner\footnote{For matrices where ill-conditioning arises mainly from diagonal elements, the point-Jacobi variant is regarded as optimal or near-optimal among matrix splitting methods\cite{templates}.}, $\mathbf{M_J}$, as
\begin{equation}\label{jacobi_precond}
    \mathbf{M_J}= Diag(\mathbf{G}).
\end{equation}
Here, $Diag(\cdot)$  extracts the diagonal elements of a matrix. 

As shown in \eqref{pEMTP}, the Jacobi preconditioner performs row-wise normalization by dividing each row by its diagonal element, essentially a form of linear scaling\cite{Jacobi}. Consequently, it cannot handle structural ill-conditioning from switching-induced row dependencies. Fig. \ref{fig:precond} illustrates this process based on the Gershgorin circle theorem.
% This letter presents \textbf{the first} attempt to address the ill-conditioning issue in EMT simulations by analyzing the network structure and examining eigenvalue distributions using the Gershgorin circle theorem.

\begin{remark}
    By demonstrating why the Jacobi preconditioner fails to improve the system matrix condition number, we validate the prior identification of switching-induced factors as the dominant source of ill-conditioning. This inevitably calls for the design of more effective preconditioners.
\end{remark}

\begin{figure*}[!htbp]%使用figure环境
\vspace{-0.4cm}
  \begin{center}
    \includegraphics[scale=0.42]{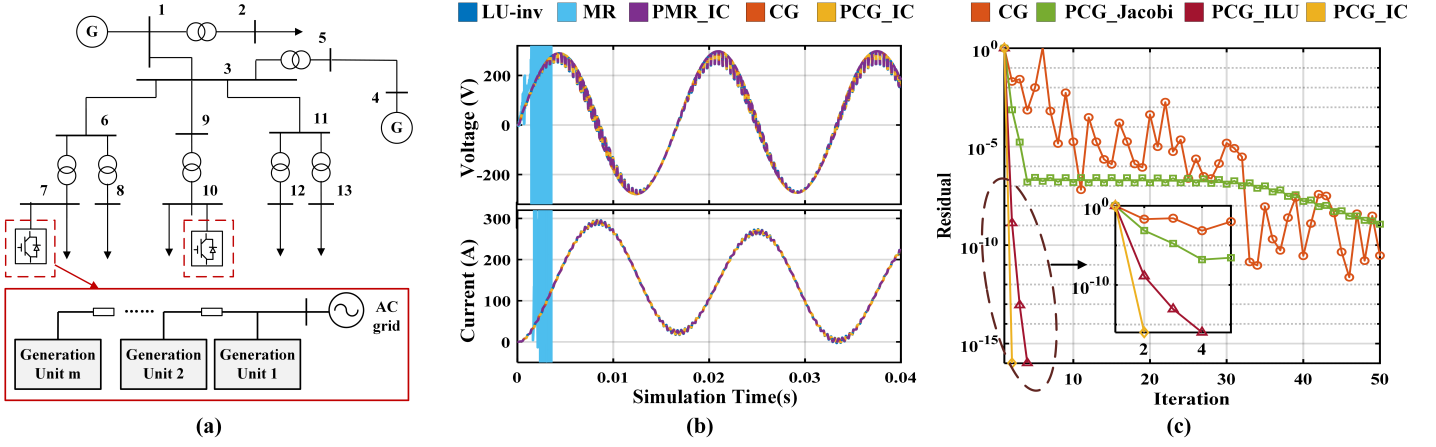}
    \setlength{\abovecaptionskip}{-0pt}
    \vspace{-0.5cm}
    \caption{Simulation results and topology of the modified IEEE 13-Bus Power System for (a) Simulation topology of the modified IEEE 13-bus power system; (b) Simulation results of phase A at the Grid-Connection point of node 7; and {(c) Iteration process at the initial time step of the EMT simulation under the nominal operating condition.}
}%添加标题
    \label{fig:topo}
  \end{center}
  \vspace{-0.6cm}
\end{figure*}

\subsubsection{Structure-informed Preconditioner for EMTP}
 Leveraging the sparsity, symmetry, and positive definiteness of the admittance matrix $\mathbf{G}$, we develop an efficient preconditioner based on the incomplete factorization (IF) method, tailored explicitly for EMTP applications. It applies fill-in rules \cite{templates} to preserve the matrix structure and directly address ill-conditioned components. Then, it can effectively alleviate the ill-conditioning inherent in EMTP problems.

Incomplete LU decomposition (ILU) is an IF method for handling sparse matrices. Its core idea is to preserve the sparsity pattern of the original matrix by restricting fill-ins to the positions of existing nonzero elements. Specifically, we construct the set as
\begin{equation}
    \mathcal{P}_{G} \triangleq\left\{(i, j): g_{i j}=0, i \neq j\right\}.
\end{equation}

When performing LU decomposition of $\mathbf{G}$, by discarding all nonzero elements at positions $(i,j)$ that are not in $\mathcal{P}_{G}$ , we obtain the zero-fill ILU factorization as

\begin{equation}\label{ILU}
    \mathbf{M_{ILU}} =\hat{L}\hat{U}=\mathbf{G}+R.
\end{equation}
Here, $\hat{L}$ and $\hat{U}$ denote the lower and upper triangular matrices obtained from the ILU factorization, $R$ is the error matrix:
\begin{equation}
    r_{i j}=\left\{\begin{array}{ll}
0, & (i, j) \notin \mathcal{P}_{G}, \\
\sum_{k=1}^{i} \hat{l}_{i k} \hat{u}_{k j}, & (i, j) \in \mathcal{P}_{G} .
\end{array}\right.
\end{equation}
Here, $r$, $\hat{l}$ and $\hat{u}$ represent the elements of the matrices $R$, $\hat{L}$ and $\hat{U}$ corresponding to specific rows and columns, respectively. Since it incorporates all ill-conditioning factors, the ILU method effectively addresses the critical structural ill-conditioning.

Building upon this, the inherent symmetry of the matrix can be further exploited through the Incomplete Cholesky (IC) factorization. This method offers improved performance and lower computing cost by constructing an approximate factorization of the form:
\begin{equation}\label{IC}
    \mathbf{M_{IC}} = \tilde{L}\tilde{L}^{T}.
\end{equation}
Here, $\tilde{L}$ denotes the lower triangular matrices obtained from the IC factorization. Based on Cholesky decomposition\cite{cholesky}, the construction of $\tilde{L}$ is as follows:
\begin{subequations}
\begin{align}
\tilde{l}_{i i}&=\sqrt{\left(g_{i i}-\sum_{k=1}^{i-1} \tilde{l}_{i k}^{2}\right)},\ \  i \in (1,n), \\
\tilde{l}_{j i}&=\frac{g_{j i}-\sum_{k=1}^{i-1} \tilde{l}_{i k} \tilde{l}_{j k}}{\tilde{l}_{i i}},\ \  j \in (i+1,n) .
\end{align}
\end{subequations}
Here, $\tilde{l}$ refers to an element in the lower triangular matrix 
$\tilde{L}$. IC preconditioner requires only half the computational effort compared to ILU, while achieving even better performance. Therefore, within the PEMTP framework, the IC preconditioner is considered the optimal solution.

Till now, we have finished the presentation of the proposed preconditioner-based method for acceleration EMTP.

\vspace{-0.2cm}
\section{Case study}
The effectiveness of the proposed method is validated on a modified IEEE 13-node system, where Nodes 7 and 10 are connected to a photovoltaic power plant comprising 50 generation units. The system includes $839$ nodes, with its detailed topology shown in Fig.\ref{fig:topo}(a). Each generation unit is modeled as a three-phase bridge converter with a voltage on the side of the DC of $900V$, resistance of 0.5$\Omega$, switching frequency of $2500Hz$, DC capacitance of $8mF$, and inductance of $1mH$. To validate the model’s accuracy comprehensively, open-loop simulations are performed. To introduce slight differences among converters, a random perturbation $5\%$ is added to the parameters of each unit.

To accurately capture the transient behavior during switching events, each converter employs a binary resistive switching model, where the conductances of the ON and OFF states are set to $1\times10^{9}$~S and $1\times10^{-9}$~S, respectively. A simulation time step of $1~\mu$s is used, which is sufficiently small to reflect the instantaneous electromagnetic transients associated with converter switching. All simulations are implemented on a Python-based EMTP solver platform.

The inherent symmetry of the matrix of the EMTP system makes CG and MR natural choices, since both are designed for symmetric matrices, and CG also requires positive definiteness. This choice ensures efficient convergence while avoiding the computational overhead associated with general nonsymmetric solvers, such as BiCGSTAB\cite{BICGSTAB}. A stringent convergence threshold of $\mathcal{E}_{set}=1\times10^{-15}$ is imposed. {The direct solver uses standard LU factorization followed by forward–backward substitution, while the KLU solver employs block triangular transformation, node reordering, and blockwise LU factorization~\cite{KLU2}.}

\begin{figure}[!htbp]%使用figure环境
\vspace{-0.3cm}
  \begin{center}
    \includegraphics[scale=0.42]{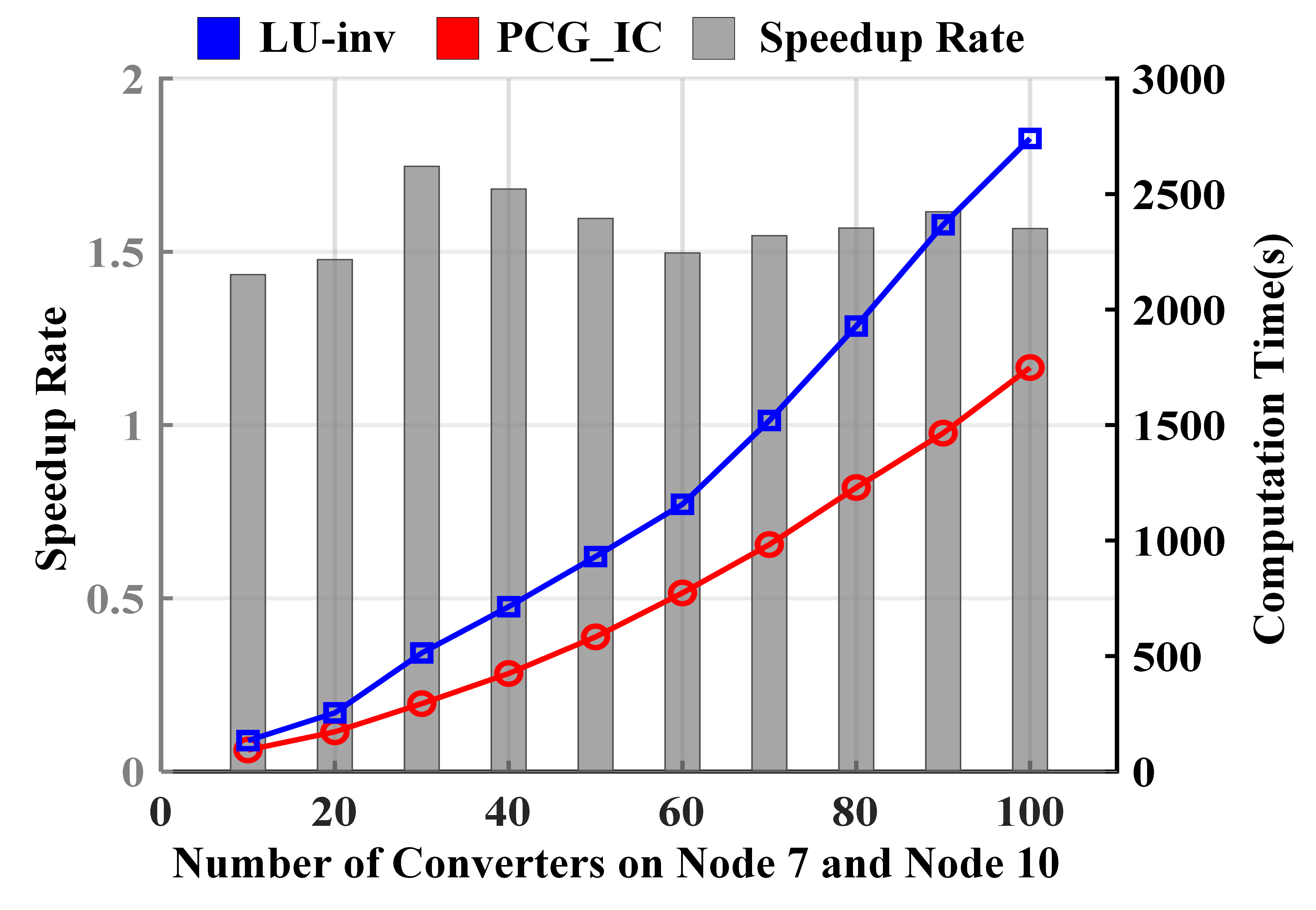}
    \setlength{\abovecaptionskip}{-0pt}
    \vspace{-0.2cm}
    \caption{Speedup factor and runtime comparison between the PCG-IC and LU methods under varying numbers of series-connected converters at Nodes 7 and 10.
}%添加标题
    \label{fig:SpeedUp}
  \end{center}
   \vspace{-0.4cm}
\end{figure}

Simulation results in Fig.~\ref{fig:topo}(b) show that the MR method fails to converge, leading to divergence and underscoring the effects of the ill-conditioned admittance matrix. In contrast, the preconditioned MR (PMR) stabilizes the iterations and achieves consistently high accuracy, with errors below $1\times10^{-10}$. Fig.~\ref{fig:topo}(c) further illustrates that all three preconditioners enhance CG convergence. Notably, the IC-based preconditioned CG (PCG-IC) converges in a single iteration, highlighting its superior efficiency and robustness. These results confirm that properly designed preconditioners can ensure both fast convergence and high accuracy in EMTP simulations, even for large-scale, severely ill-conditioned systems.

Table \ref{table.time} presents a detailed performance comparison. Without preconditioning, the CG method is slower than direct inversion, while MR offers marginal speed gains but suffers from poor convergence due to matrix ill-conditioning, resulting in significant numerical errors. Introducing preconditioners greatly enhances both accuracy and convergence. 
{Among all tested preconditioners, the IC-based approach delivers the best overall performance. Fig.~\ref{fig:SpeedUp} reports the detailed runtime comparisons and acceleration factors across different network sizes. As the number of converters increases from N=10 (199 nodes) to N=100 (1639 nodes), corresponding to more than an eightfold increase in network size, the proposed method consistently maintains a speedup above 1.5×, approaching twofold acceleration at moderate scales, while preserving high numerical accuracy with errors below $1\times10^{-10}$. These results demonstrate the strong scalability and robustness of the IC-preconditioned solver for large-scale EMT simulations.}
%Among them, the IC preconditioner achieves the best performance, achieving nearly twofold acceleration and maintaining over 1.5× speedup even at $N=100$, with numerical errors below $1\times10^{-10}$. The detailed runtime comparison and acceleration factors for converter counts ranging from $N=10$ to $100$ are illustrated in Fig.~\ref{fig:SpeedUp}, demonstrating the stable and scalable efficiency of the proposed approach.

It is noteworthy that the proposed PCG-IC method significantly outperforms the widely used KLU solver\cite{KLU}.This because KLU does not mitigate the matrix's intrinsic ill-conditioning, these steps become computationally intensive. In contrast, the IC preconditioner directly mitigates the sources of ill-conditioning, transforming $\mathbf{G}$ into a well-conditioned, near-identity form. As shown in Table~\ref{table.time}, the preconditioned matrix achieves a condition number close to unity, and PCG converges in an average of only 1.023 iterations, demonstrating superior efficiency for large-scale EMTP systems.

\begin{figure}[!htbp]%使用figure环境
\vspace{-0.5cm}
  \begin{center}
    \includegraphics[scale=0.5]{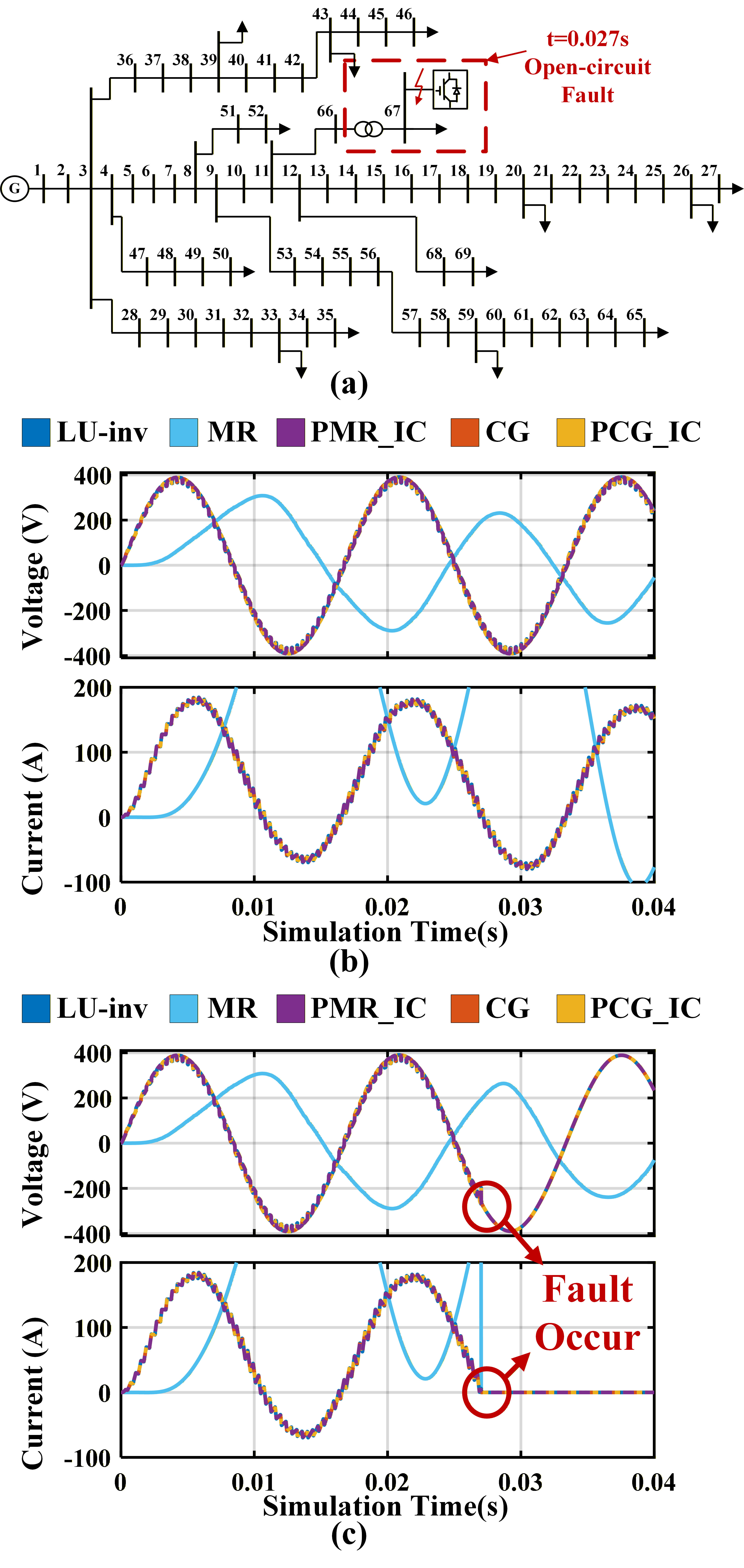}
    \setlength{\abovecaptionskip}{-0pt}
    \vspace{-0.2cm}
    \caption{{Simulation results and topology of the modified IEEE 69-bus system: 
(a) Topology of the modified IEEE 69-bus network; 
(b) steady-state simulation results of phase A at the grid-connection point of Node 67; and 
(c) transient simulation results following an open-circuit fault at $t = 0.027$ s on the connection line between the photovoltaic plant and Node 67.}
}%添加标题
    \label{fig:IEEE69}
  \end{center}
   \vspace{-0.6cm}
\end{figure}

{
To further assess scalability, the proposed method is evaluated on a modified IEEE 69-bus system with a photovoltaic plant consisting of 100 generation units connected at Node 67, resulting in a network with over 1,000 nodes. The photovoltaic parameters remain identical to those used previously, and the system topology and results are shown in Fig. \ref{fig:IEEE69}.}

\begin{table}[!htbp]
\vspace{-0.3cm}
\renewcommand{\arraystretch}{1.2}
\caption{ Method Comparison on the Modified IEEE 13-Bus System}
\vspace{-0.2cm}
\label{table.time}
\centering
% \usepackage{multirow}
\begin{tabular}{c|c|c|c|c|c}
\toprule
\hline
\begin{tabular}[c]{@{}c@{}}Precondi-\\ tioner\end{tabular} & \begin{tabular}[c]{@{}c@{}}Condition\\ Number\end{tabular} & Method & Iteration & Time(s) & { \begin{tabular}[c]{@{}c@{}}RMSE\\ ($10^{-10}$)\end{tabular}} \\ \hline
 & { } & LU-inv &  & 957.130 & { 0.00925} \\ \cline{3-3} \cline{5-6} 
 & { } & KLU & \multirow{-2}{*}{/} & 607.655 & { 0.10570} \\ \cline{3-6} 
 & { } & MR & 170.3 & 882.934 & { \textbf{Diverge}} \\ \cline{3-6} 
\multirow{-4}{*}{None} & \multirow{-4}{*}{{ $6.502\times10^{12}$}} & CG & 769.6 & 1628.51 & { 0.15701} \\ \hline
 & { } & PMR & 115.7 & 771.882 & { 0.28804} \\ \cline{3-6} 
\multirow{-2}{*}{Jacobi} & \multirow{-2}{*}{{ $6.502\times10^{12}$}} & PCG & 136.7 & 721.959 & { 0.06573} \\ \hline
 &  & PMR & 5.951 & 634.107 & { 1.87783} \\ \cline{3-6} 
\multirow{-2}{*}{ILU} & \multirow{-2}{*}{4.3927} & PCG & 5.023 & 613.542 & { 0.01494} \\ \hline
 &  & PMR & 1.337 & 598.013 & { 0.01331} \\ \cline{3-6} 
\multirow{-2}{*}{\textbf{IC}} & \multirow{-2}{*}{\textbf{1.0000}} & \textbf{PCG} & \textbf{1.023} & \textbf{579.486} & { \textbf{0.01299}} \\ \hline
\end{tabular}
% \begin{tabular}{c|c|c|c|c}
% \toprule
% \hline
% \begin{tabular}[c]{@{}c@{}}Precondi-\\ tioner\end{tabular} & \begin{tabular}[c]{@{}c@{}}condition\\ number\end{tabular} & Method & \begin{tabular}[c]{@{}c@{}}simulation\\  time(s)\end{tabular} & RMSE \\ \hline
% \multirow{3}{*}{None} & \multirow{3}{*}{323070145} & LU-inv & 230.618731 & 0 \\ \cline{3-5} 
%  &  & GMRES & 1359.33560 & 1 \\ \cline{3-5} 
%  &  & CG & 202.449741 & 1 \\ \hline
% \multirow{2}{*}{Jacobi} & \multirow{2}{*}{140.015} & PGMRES & 215.860278 & 1 \\ \cline{3-5} 
%  &  & \textbf{PCG} & \textbf{119.541342} & 1 \\ \hline
% \multirow{2}{*}{ILU} & \multirow{2}{*}{4.58079} & PGMRES & 130.805663 & 1 \\ \cline{3-5} 
%  &  & PCG & 128.741656 & 1 \\ \hline
% \multirow{2}{*}{IC} & \multirow{2}{*}{1.00000} & PGMRES & 124.973941 & 1 \\ \cline{3-5} 
%  &  & PCG & 121.352320 & 1 \\ \hline
% \end{tabular}
\vspace{-0.4cm}
\end{table}

{
Figure \ref{fig:IEEE69}(b) shows the steady-state simulation results obtained with different solvers and preconditioners. Without preconditioning, the MR method exhibits noticeable deviations, and the Jacobi preconditioner fails to sufficiently improve the numerical accuracy, with voltage errors remaining on the order of $10^{-2}$. In contrast, both ILU and IC preconditioners effectively mitigate the ill-conditioning of the EMT system matrix, reducing the error to approximately $10^{-7}$. Among all tested approaches, the PCG-IC method achieves the best overall performance.}

{
Similar trends are observed in the transient simulations shown in Fig. \ref{fig:IEEE69}(c). Following an open-circuit fault at $t=0.027$ s between the photovoltaic plant and Node 67, the MR method again suffers from significant numerical errors, whereas the PCG-IC method maintains stable numerical behavior, delivering consistently high accuracy and the shortest overall simulation time.
This consistency with the modified IEEE 13-bus results further suggests that the proposed method maintains stable acceleration performance across different network complexities and transient scenarios.}

{
Table \ref{table.time2} further compares different preconditioning strategies. Although algebraic multigrid (AMG) preconditioning improves convergence and accuracy, its high setup and smoothing costs lead to substantially lower efficiency in EMT simulations, where linear systems must be solved repeatedly at each time step.}

\vspace{-0.5cm}
\section{{Discussion}}
{
In EMT simulations, the system admittance matrix is symmetric in most practical cases, since it represents the instantaneous physical network, and control actions affect it only through updates of switching states or equivalent parameters, rather than through explicit coupling terms in the nodal equations. Mutual coupling typically appears in transformer models and is commonly represented by T- or $\pi$-type equivalent circuits, which preserve reciprocity and thus maintain symmetry. The main source of asymmetry arises from voltage-controlled sources, whose Norton equivalents introduce voltage-dependent current injections. However, such elements are rarely encountered in power system EMT studies; in these cases, nonsymmetric solvers (e.g., BiCGSTAB or Quasi-Minimal Residual) combined with ILU-type preconditioners are more suitable.}

{
Regarding positive definiteness, practical power networks almost always include grounding paths (e.g., source grounding or shunt-to-ground admittances). Consequently, the corresponding admittance matrices are typically irreducible weakly diagonally dominant, which ensures positive definiteness. When indefiniteness does arise, MR-type methods, which do not require positive definiteness, become preferable.}

\section{Conclusions}\label{conclu}
%This letter presents the first mathematical acceleration of EMTP linear equation solving using preconditioner-based methods. By analyzing mathematical properties derived from the system’s physical characteristics, it bridges the gap between physical modeling and numerical algorithms, offering new insights for EMTP acceleration.

{This letter introduces the first systematic preconditioner-base acceleration of EMTP linear solvers through a systematic mitigation of the underlying matrix ill-conditioning.} Our core contribution is to reveal the pathological sources of ill-conditioning within EMTP networks and systematically link them to physical mechanisms such as switching devices, thereby interpreting numerical difficulties as intrinsic consequences of the system’s physical properties rather than as isolated computational issues. This analysis serves not only as the foundation for our preconditioner-based acceleration strategy but also as a powerful diagnostic tool for identifying and quantifying the numerical risks posed by specific network topologies or component parameters. The proposed analysis bridges the gap between physical modeling and numerical algorithms, opening a new avenue for EMTP acceleration and establishing a theoretical foundation for future physics-informed solvers.

\begin{table}[!htbp]
\vspace{-0.5cm}
\renewcommand{\arraystretch}{1.2}
\caption{{Method Comparison on the Modified IEEE 69-Bus System}}
\vspace{-0.2cm}
\label{table.time2}
\centering
\begin{tabular}{c|c|cc|cc}
\toprule
\hline
{ } & { } & \multicolumn{2}{c|}{{ Steady-state}} & \multicolumn{2}{c}{{ Transient}} \\ \cline{3-6} 
\multirow{-2}{*}{{ \begin{tabular}[c]{@{}c@{}}Precondi-\\ tioner\end{tabular}}} & \multirow{-2}{*}{{ Method}} & \multicolumn{1}{c|}{{ Time(s)}} & { \begin{tabular}[c]{@{}c@{}}RMSE\\ (1e-6)\end{tabular}} & \multicolumn{1}{c|}{{ Time(s)}} & { \begin{tabular}[c]{@{}c@{}}RMSE\\ (1e-6)\end{tabular}} \\ \hline
{ } & { LU-inv} & \multicolumn{1}{c|}{{ 1422.585}} & { 0.10561} & \multicolumn{1}{c|}{{ 1435.607}} & { 0.10573} \\ \cline{2-6} 
{ } & { KLU} & \multicolumn{1}{c|}{{ 858.741}} & { 0.12559} & \multicolumn{1}{c|}{{ 843.893}} & { 0.12600} \\ \cline{2-6} 
{ } & { MR} & \multicolumn{1}{c|}{{ 660.247}} & { 384769} & \multicolumn{1}{c|}{{ 719.375}} & { 18622.4} \\ \cline{2-6} 
\multirow{-4}{*}{{ None}} & { CG} & \multicolumn{1}{c|}{{ 28334.6}} & { 3.15830} & \multicolumn{1}{c|}{{ 18185.6}} & { 3.53283} \\ \hline
{ } & { PMR} & \multicolumn{1}{c|}{{ 952.976}} & { 4956.31} & \multicolumn{1}{c|}{{ 1117.38}} & { 4432.17} \\ \cline{2-6} 
\multirow{-2}{*}{{ Jacobi}} & { PCG} & \multicolumn{1}{c|}{{ 29628.4}} & { 0.89604} & \multicolumn{1}{c|}{{ 1621.38}} & { 0.93021} \\ \hline
{ } & { PMR} & \multicolumn{1}{c|}{{ 824.516}} & { 422.468} & \multicolumn{1}{c|}{{ 837.432}} & { 426.399} \\ \cline{2-6} 
\multirow{-2}{*}{{ ILU}} & { PCG} & \multicolumn{1}{c|}{{ 827.969}} & { 0.13625} & \multicolumn{1}{c|}{{ 832.939}} & { 0.13478} \\ \hline
{ } & { PMR} & \multicolumn{1}{c|}{{ 817.091}} & { 1.59049} & \multicolumn{1}{c|}{{ 810.124}} & { 1.59255} \\ \cline{2-6} 
\multirow{-2}{*}{{ \textbf{IC}}} & { \textbf{PCG}} & \multicolumn{1}{c|}{{ \textbf{803.187}}} & { \textbf{0.13022}} & \multicolumn{1}{c|}{{ \textbf{809.806}}} & { \textbf{0.12954}} \\ \hline
{ } & { PMR} & \multicolumn{1}{c|}{{ 59259.1}} & { 30.6790} & \multicolumn{1}{c|}{{ 59843.9}} & { 26.5742} \\ \cline{2-6} 
\multirow{-2}{*}{{ AMG}} & { PCG} & \multicolumn{1}{c|}{{ 81826.1}} & { 0.27228} & \multicolumn{1}{c|}{{ 82048.8}} & { 0.26559} \\ \hline
\end{tabular}
\vspace{-0.5cm}
\end{table}

% \section*{Acknowledgment}
% This work was supported, in part, by the U.S. National Science Foundation under Grant 1917308 and by the United States Department of Energy Office of Electricity Advanced system Modeling Program, and performed under the auspices of the U.S. Department of Energy by Lawrence Livermore National Laboratory under Contract DE-AC52-07NA27344. Document released as LLNL-PROC-XXXXXX.

 \newcommand{\BIBdecl}{\setlength{\itemsep}{0.01 em}}
 \bibliographystyle{IEEEtran}
 \vspace{-0.2cm}
\bibliography{IEEEabrv,Reference.bib}

@IEEEtranBSTCTL{IEEEexample:BSTcontrol,
CTLuse_forced_etal       = "yes",
CTLmax_names_forced_etal = "5",
CTLdash_repeated_names = "no",
CTLnames_show_etal       = "1" }

@book{EMTPBOOK,
    author = {Dommel, H.W.},
    title = {EMTP Theory Book},
    publisher = {Microtran Power System Analysis Corporation,Vancouver, Canada},
    year = {1992}
}

@article{Gershgorin_circle,
  title={Gershgorin's theorem and the zeros of polynomials},
  author={Bell, Howard E},
  journal={The American Mathematical Monthly},
  volume={72},
  number={3},
  pages={292--295},
  year={1965},
  publisher={JSTOR}
}

@book{templates,
  title={Templates for the solution of linear systems: building blocks for iterative methods},
  author={Barrett, Richard and Berry, Michael and Chan, Tony F and Demmel, James and Donato, June and Dongarra, Jack and Eijkhout, Victor and Pozo, Roldan and Romine, Charles and Van der Vorst, Henk},
  year={1994},
  publisher={SIAM}
}

@article{KLU,
  title={Algorithm 907: KLU, a direct sparse solver for circuit simulation problems},
  author={Davis, Timothy A and Palamadai Natarajan, Ekanathan},
  journal={ACM Transactions on Mathematical Software (TOMS)},
  volume={37},
  number={3},
  pages={1--17},
  year={2010},
  publisher={ACM New York, NY, USA}
}

@book{cg_mires,
  title={Iterative methods for sparse linear systems},
  author={Saad, Yousef},
  year={2003},
  publisher={SIAM}
}

@article{subdim,
  title={Enabling large-scale and high-precision fluid simulations on near-term quantum computers},
  author={Chen, Zhao-Yun and Ma, Teng-Yang and Ye, Chuang-Chao and Xu, Liang and Bai, Wen and Zhou, Lei and Tan, Ming-Yang and Zhuang, Xi-Ning and Xu, Xiao-Fan and Wang, Yun-Jie and others},
  journal={Computer Methods in Applied Mechanics and Engineering},
  volume={432},
  pages={117428},
  year={2024},
  publisher={Elsevier}
}

@article{TPS1,
  title={A class of new preconditioners for linear solvers used in power system time-domain simulation},
  author={Khaitan, Siddhartha Kumar and McCalley, James D},
  journal={IEEE Transactions on Power Systems},
  volume={25},
  number={4},
  pages={1835--1844},
  year={2010},
  publisher={IEEE}
}

@ARTICLE{CSEE2,
  author={Wang, Chengshan and Yu, Hao and Li, Peng and Wu, Jianzhong and Ding, Chengdi and Song, Guanyu and Lin, Dun and Xing, Feng},
  journal={CSEE Journal of Power and Energy Systems}, 
  title={EMTP-type program realization of Krylov subspace based model reduction methods for large-scale active distribution network}, 
  year={2015},
  volume={1},
  number={1},
  pages={52-60},
  doi={10.17775/CSEEJPES.2015.00007}}

@ARTICLE{CSEE1,
  author={Zhang, Yi and Ding, Hui and Kuffel, Rick},
  journal={CSEE Journal of Power and Energy Systems}, 
  title={Key techniques in real time digital simulation for closed-loop testing of HVDC systems}, 
  year={2017},
  volume={3},
  number={2},
  pages={125-130},
  doi={10.17775/CSEEJPES.2017.0016}}

@article{cholesky,
  title={Analysis of the Cholesky decomposition of a semi-definite matrix},
  author={Higham, Nicholas J},
  year={1990},
  publisher={Oxford University Press}
}

@article{MMCmodel,
  title={An efficient half-bridge MMC model for EMTP-type simulation based on hybrid numerical integration},
  author={Gao, Shilin and Chen, Ying and Song, Yankan and Yu, Zhitong and Wang, Yenan},
  journal={IEEE Transactions on Power Systems},
  volume={39},
  number={1},
  pages={1162--1177},
  year={2023},
  publisher={IEEE}
}

@inproceedings{KLU2,
  title={A comparative analysis of LU decomposition methods for power system simulations},
  author={Razik, Lukas and Schumacher, Lennart and Monti, Antonello and Guironnet, Adrien and Bureau, Gautier},
  booktitle={2019 IEEE Milan PowerTech},
  pages={1--6},
  year={2019},
  organization={IEEE}
}

@article{CG,
  title={The conjugate gradient method for linear and nonlinear operator equations},
  author={Daniel, James W},
  journal={SIAM Journal on Numerical Analysis},
  volume={4},
  number={1},
  pages={10--26},
  year={1967},
  publisher={SIAM}
}

@article{MR,
  title={Least squares residuals and minimal residual methods},
  author={Liesen, J{\"o}rg and Rozlozn{\'\i}k, Miroslav and Strakos, Zdenek},
  journal={SIAM Journal on Scientific Computing},
  volume={23},
  number={5},
  pages={1503--1525},
  year={2002},
  publisher={SIAM}
}

@article{Jacobi,
  title={A modified Jacobi preconditioner for solving ill-conditioned Biot's consolidation equations using symmetric quasi-minimal residual method},
  author={Chan, SH and Phoon, KK and Lee, FH},
  journal={International Journal for Numerical and Analytical Methods in Geomechanics},
  volume={25},
  number={10},
  pages={1001--1025},
  year={2001},
  publisher={Wiley Online Library}
}

@article{BICGSTAB,
  title={Variants of BICGSTAB for matrices with complex spectrum},
  author={Gutknecht, Martin H},
  journal={SIAM journal on scientific computing},
  volume={14},
  number={5},
  pages={1020--1033},
  year={1993},
  publisher={SIAM}
}

@article{EMT_decouple,
  title={A novel decoupled EMT approach and parallel simulation framework for modularized solid-state transformers},
  author={Feng, Moke and Gao, Chenxiang and Xu, Jianzhong and Zhao, Chengyong and Li, Gen},
  journal={IEEE Transactions on Power Delivery},
  volume={38},
  number={5},
  pages={3285--3295},
  year={2023},
  publisher={IEEE}
}

@article{EMT_model,
  title={A simplified EMT model of multiple-active-bridge based power electronic transformer with integrated energy storage},
  author={Xu, Jianzhong and Zheng, Conghui and Xu, Wanying and Feng, Moke and Zhao, Chengyong and Li, Gen},
  journal={CSEE Journal of Power and Energy Systems},
  year={2024},
  publisher={CSEE}
}

@article{EMT_GPU,
  title={Real-time electromagnetic transient simulation of multi-terminal hvdc--ac grids based on gpu},
  author={Sun, Jingfan and Debnath, Suman and Saeedifard, Maryam and Marthi, Phani RV},
  journal={IEEE Transactions on Industrial Electronics},
  volume={68},
  number={8},
  pages={7002--7011},
  year={2020},
  publisher={IEEE}
}

@article{EMT_FPGA,
  title={A fixed-admittance algorithm for the FPGA-based microsecond-level nonlinear real-time simulation of the hybrid DCCB},
  author={Su, Hang and Xu, Jin and Zhou, Jianqi and Qi, Zhiping and Yu, Jianghua and Wang, Keyou},
  journal={CSEE Journal of Power and Energy Systems},
  year={2025},
  publisher={CSEE}
}

\begin{IEEEbiography}
[{\includegraphics[width=1in,height=1.25in,clip,keepaspectratio]{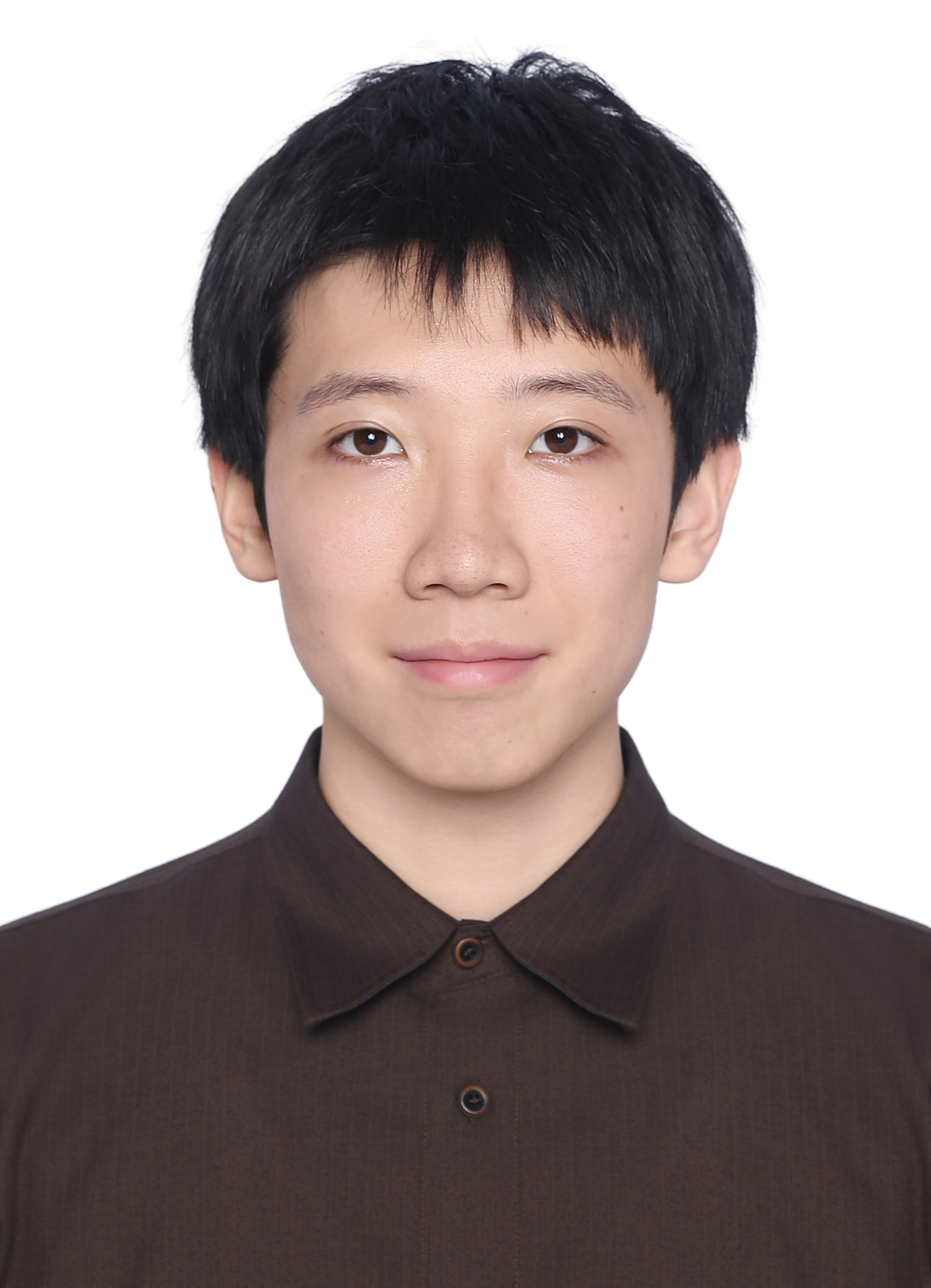}}]{Qi Lou}(Student Member, IEEE) received the B.S. degree in Electrical Engineering from Southeast University, Nanjing, China, in 2024. He is currently working toward the Ph.D. degree in Electrical Engineering from Southeast University, Nanjing, China. His research interests include quantum electromagnetic transient simulation.
\end{IEEEbiography}

\begin{IEEEbiography}[{\includegraphics[width=1in,height=1.25in,clip,keepaspectratio]{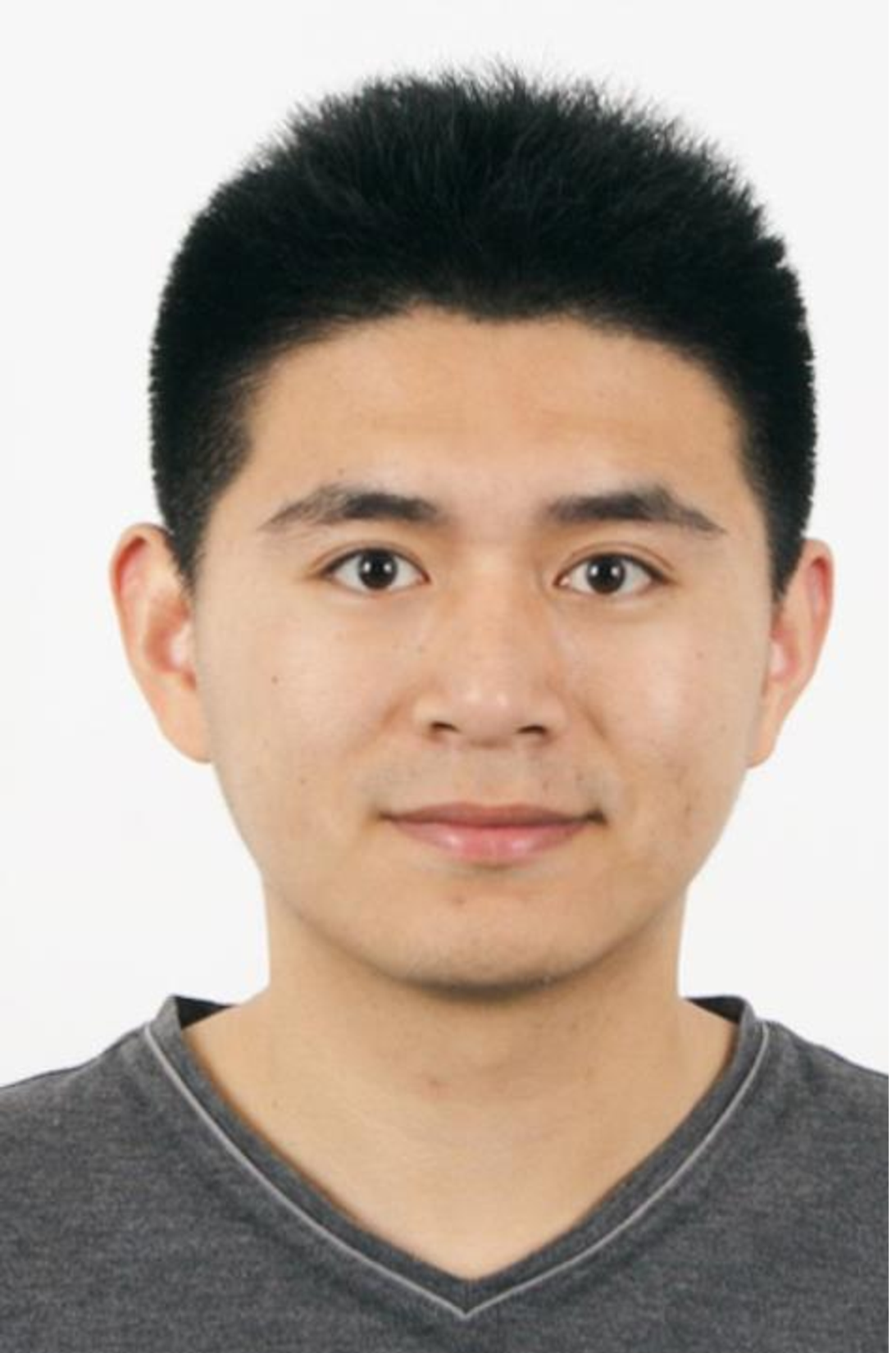}}]{Yijun Xu}(Senior Member, IEEE) is a professor at Southeast University, Nanjing, China. He received his Ph.D. degree from the Bradley Department of Electrical and Computer Engineering at Virginia Tech, Falls Church, VA, in December 2018. He worked as a research assistant professor at Virginia Tech-Northern Virginia Center, Falls Church, VA, in 2021. He was a postdoc associate at the same institute from 2019 to 2020.  
He did a computation internship at Lawrence Livermore  National Laboratory, Livermore, CA, and a power engineer internship at ETAP -- Operation Technology, Inc., Irvine, California, in 2018 and 2015, respectively. 

His research interests include power system uncertainty quantification, uncertainty inversion, and decision-making under uncertainty. 
Dr. Xu is currently serving as an Associate Editor of the \textsc{IET Generation, Transmission \& distribution}, an Associate Editor of the \textsc{IET Renewable Power Generation}, and the Young Editor of the \textsc{Power System Protection and Control}. He is the co-chair of the IEEE Task Force on Power System Uncertainty Quantification and Uncertainty-Aware Decision-Making.
\end{IEEEbiography}

\begin{IEEEbiography}[{\includegraphics[width=1in,height=1.25in,clip,keepaspectratio]{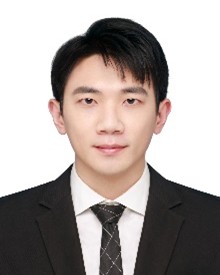}}]{Yang Cao} (Student Member, IEEE) received the B.S. and M.S. degrees in Power Engineering from Southeast University, Nanjing, China, in 2018 and 2021. He is currently working toward the Ph.D. degree in Electrical Engineering from Southeast University, Nanjing, China. 
His research interests include modeling, control, and real-time simulation of power electronic systems.
\end{IEEEbiography}

\begin{IEEEbiography}[{\includegraphics[width=1in,height=1.25in,clip,keepaspectratio]{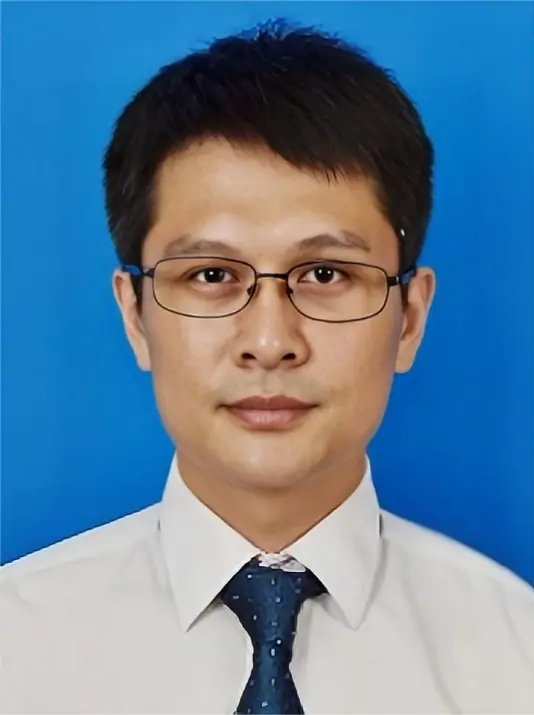}}]{Wei Gu}(Senior Member, IEEE) received his B.S. and Ph.D. degrees in Electrical Engineering from Southeast University, China, in 2001 and 2006, respectively. From 2009 to 2010, he was a Visiting Scholar in the Department of Electrical Engineering, Arizona State University. 

He is now a professor at the School of Electrical Engineering, Southeast University. He is the director of the institute of distributed generations and active distribution networks. His research interests include distributed generations and microgrids, integrated energy systems. He is an Editor for the IEEE Transactions on Power Systems, the IET Energy Systems Integration and the Automation of Electric Power Systems (China). 
\end{IEEEbiography}

\begin{IEEEbiography}[{\includegraphics[width=1in,height=1.25in,clip,keepaspectratio]{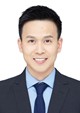}}]{Fei Zhang}(Member, IEEE) received the B.S. and M.S. degrees in electrical engineering from Tsinghua University, Beijing, China, in 2009 and 2012, respectively, and the Ph.D. degree in electrical engineering from McGill University, Montreal, Canada, in 2018. From 2018 to 2020, he was a specialist in modeling and electrical simulation with Opal-RT Technologies, Montreal, Canada. Since 2020, he has been an associate professor with school of electrical engineering, Southeast University, Nanjing, China. His research interest includes HVDC converters, high power electronics, and real-time simulation.
\end{IEEEbiography}

\vspace{-0.3cm}
\end{document}